\documentclass[10pt,twocolumn]{article}

\usepackage[margin=1in, columnsep=0.2in]{geometry}
\usepackage{multicol}
\usepackage[font=small,labelfont=bf,justification=justified,width=\textwidth,skip=8pt]{caption}
\usepackage{placeins}
\usepackage{afterpage}
\usepackage{balance}  
\usepackage{dblfloatfix}  
\usepackage{flushend}  
\usepackage{graphicx}
\usepackage{amsmath}
\usepackage{amsfonts}
\usepackage{amssymb}
\usepackage{bm}
\usepackage{units}
\usepackage{yfonts}
\usepackage{subfigure}
\usepackage[table]{xcolor}
\usepackage{soul}
\usepackage{hyperref}
\usepackage{svg}
\usepackage{float}
\usepackage[normalem]{ulem}
\usepackage{authblk}
\usepackage{setspace}
\usepackage[scaled]{helvet}
\usepackage{braket}
\usepackage{multirow}
\usepackage{mathrsfs}
\usepackage{booktabs}
\usepackage{algorithm}
\usepackage{algorithmicx}
\usepackage{algpseudocode}
\usepackage{listings}
\usepackage{xcolor}
\usepackage[utf8]{inputenc}
\usepackage{textcomp}
\usepackage{newunicodechar}
\newunicodechar{−}{-}

\newcommand{\fighel}{\sffamily}

\title{Acoustic Metamaterials with Positive and Negative Couplings: Modular and One Piece Architectures for Topological Models}

\author{Jackson Saunders\textsuperscript{[1,*]}, Camelia Prodan\textsuperscript{1}}

\date{} 

\renewenvironment{abstract}%
{\small\begin{center}\textbf{Abstract}\end{center}\begin{quotation}\small}%
{\end{quotation}}

\usepackage{titlesec}
\titleformat{\section}{\large\bfseries}{\thesection}{1em}{}
\titleformat{\subsection}{\normalsize\bfseries}{\thesubsection}{1em}{}
\titlespacing*{\section}{0pt}{12pt plus 4pt minus 2pt}{6pt plus 2pt minus 2pt}
\titlespacing*{\subsection}{0pt}{8pt plus 2pt minus 2pt}{4pt plus 2pt minus 2pt}

\begin{document}

\twocolumn[
\begin{@twocolumnfalse}
\maketitle

\begin{center}
\small
\textsuperscript{1} Department of Physics and Engineering Physics, Fordham University, 441 East Fordham Rd., Bronx, NY 10458, USA.\\

\bigskip
\textsuperscript{*}Correspondence: \href{mailto:Jsaunders26@fordham.edu}{Jsaunders26@fordham.edu}
\end{center}

\begin{abstract}
We describe two 3D-printing approaches for realizing tight-binding models in acoustic metamaterials using  H shaped-resonators: a modular system with tunable interconnections and an integrated one piece design for reducing dissipation. The platform supports both positive and negative coupling through geometric control, enabling accurate acoustic analogs of topological models. By tuning the coupling length (CL), we eliminate detuning effects and preserve particle–hole symmetry. We further quantify the influence of the Total Coupling Area (TCA) on band topology and derive conditions for constant-area coupling. The system was tested on SSH and Kitaev chains revealing midgap edge and interface states, confirming topological behavior in both configurations.


\end{abstract}

\vspace{0.5cm}
\end{@twocolumnfalse}]


\section{Introduction}

For the past two decades, acoustics has presented a viable platform for the realization of topological phenomena \cite{yang2015topological,xue2022topological}. The ability to engineer acoustic metamaterials with precisely controlled geometries and coupling strengths has enabled the exploration of condensed matter physics concepts in classical wave systems, offering unique advantages such as direct visualization of wave functions and tunable parameters \cite{ma2019topological,zhang2018topological}. These acoustic analogs have proven particularly valuable for studying topological insulators, where edge states emerge from bulk-boundary correspondence, and more exotic phases of matter that are challenging to realize in their original electronic contexts \cite{huber2016topological}.
\begin{figure*}[!t]
    \def\svgwidth{\textwidth}
    \centering
    {\fighel
        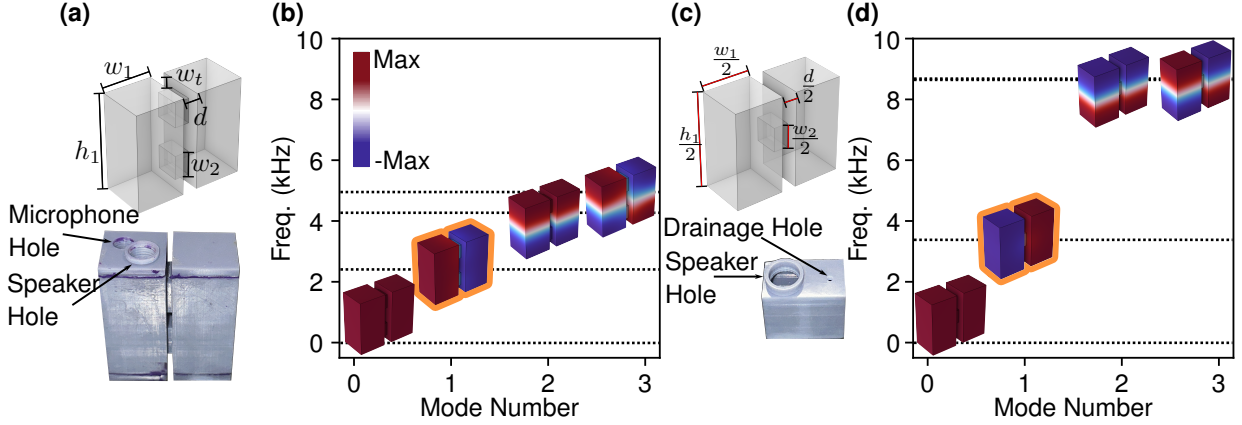
        }
    \caption{\textbf{H-resonator design and spectra.} \textbf{(a)} Cad model of double (top) and single (bottom) bridge H-resonator with dimensions $h_1 = 40$ mm, $w_1 = 20$ mm, $wt = 4$  mm, $d = 6$ mm, \& $w_2 = 10$ mm. \textbf{(b)} To-scale 3D printed resonator comparison. Larger double-bridge resonator (top) printed hollow then sealed with a cap using puddy. Smaller single-bridge resonator (bottom) printed fully encased. \textbf{(c)} Resonance spectrum and corresponding modes for the double-bridge resonator. Pressure legend inlayed in the top right. \textbf{(d)} Resonance spectrum and corresponding modes for the single-bridge resonator. \textbf{(c)} \& \textbf{(d)} Selected $f_o$ mode that \textit{all} later systems are designed around are highlighted in green. For the modular systems $f_o^m = 2.38$ kHz and for the single-shot systems $f_o^s = 3.38$ kHz}
    \label{fig:h-res-design}
\end{figure*}

\begin{figure*}[!t]
    \def\svgwidth{\textwidth}
    \centering
    {\fighel
        \input{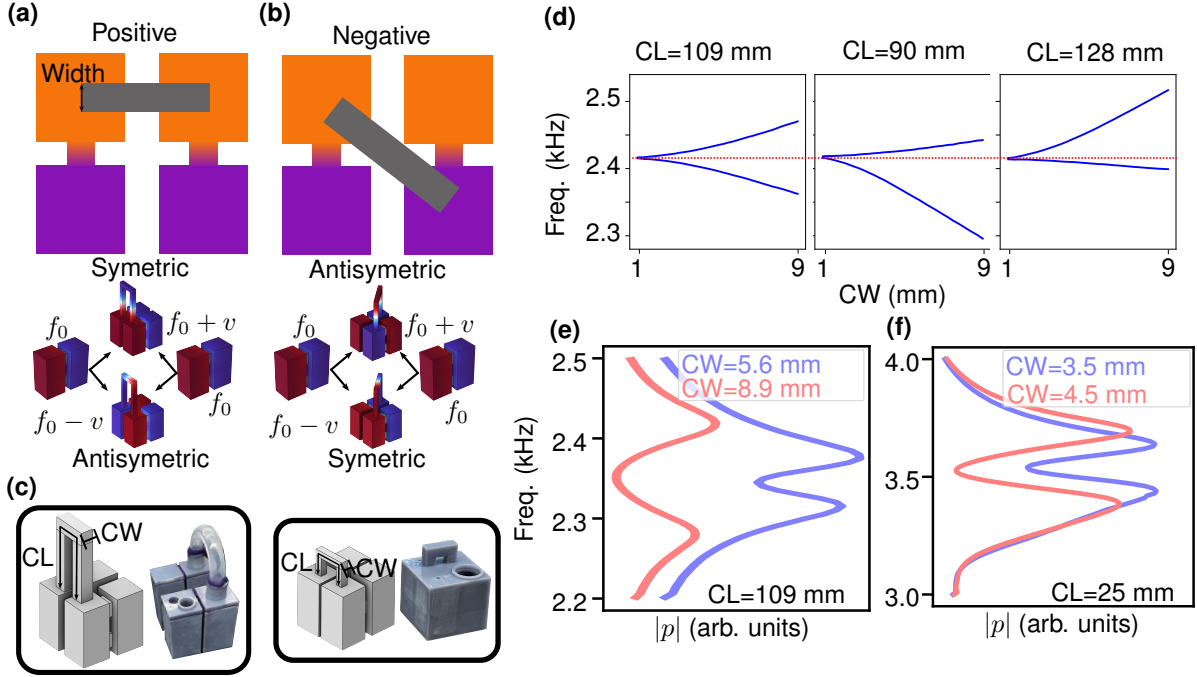}
    }
    \caption{\textbf{Positive/negative coupling schematic and hybridization diagrams with simulated and experimental dimer results.} \textbf{(a)} Internal reference frame of resonators denoted with orange and purple. Positive coupling is defined as coupling orange to orange. Negative coupling is defined as coupling Purple to Orange. \textbf{(b)} Hybridization diagrams for the H-resonator describing the symmetric and antisymmetric modes corresponding to positive (left) and negative (right) couplings corresponding to $H_C^+$ and $H_C^-$ respectively Eq.\ref{eq:2} \textbf{(c)} CAD and experimental set up for both double (left) and single (right) bridge dimer apparatus. \textbf{(d)} Resonance spectra for the dimer as a function of CW. Each plot represents a different geometry defined by a unique CL value. Symmetric plot corresponding to $\epsilon = 0$ (left) and asymmetric plot corresponding to $\epsilon \ne 0$ (center \& right). \textbf{(e)} Experimental measurements of dimer apparatus demonstrating symmetry at various coupling widths for the modular system with CL = 109 mm. \textbf{(f)} Experimental measurements of dimer apparatus demonstrating symmetry at various coupling widths for single-shot dimer CL=25mm.}
    \label{fig:dimer-panel}
\end{figure*}

Recent advances in topological acoustics have bifurcated into two architectures. On one side are diffraction/propagation platforms that rely on waveguide networks \cite{latice1,latice2, latice3}. On the other side are cavity-based, locally resonant metamaterials that implement tight-binding physics with printed, individual resonators and discrete, tube-mediated couplings \cite{cavity1, cavity2, cavity3, cavity4, cavity5, cavity6, cavity7, cavity8}. The latter route, as described here, offers direct control of hopping sign and magnitude, reduced sensitivity to environmental noise, and straightforward scalability/reconfiguration. These developments collectively motivate a cavity-based approach for programmable tight-binding models where sign-alternating couplings and on-site terms can be tuned independently, which is essential for models like the one explored here.

While this method has advanced significantly, the increasing complexity of models has imposed additional constraints on their design and manufacturing. For instance, theoretical models have greatly increased the required number of couplings between resonators. This can make it difficult to route and manufacture the couplings. Additionally, the limited surface area of the individual resonators and the requirement that the couplings be connected from the top or bottom\cite{chiralsym} can require one to make larger resonators; increasing manufacturing time and cost.

In this work, we address these challenges by developing two complementary manufacturing methods for cavity based acoustic metamaterials based on H-resonators connected via tunable tube couplings. The versatility of the tube based coupling can circumvent issues imposed by spacial limitations while the H-resonator design, consisting of two coupled chambers, provides a more coupling area. Crucially, our tube-based coupling scheme enables the implementation of both positive and negative hopping amplitudes by controlling the connection geometry between resonators. This capability is essential for realizing models with alternating coupling signs, such as the Kitaev model of a 1-D topological superconductor\cite{Kitaev2001}.

Our first approach employs a modular system where individual H-resonators are 3D-printed with interchangeable caps, allowing for rapid prototyping and systematic parameter exploration through reconfigurable tube connections. The second approach utilizes single-shot 3D printing to fabricate entire metamaterial arrays as monolithic structures, providing enhanced mechanical stability and eliminating potential acoustic losses at connection interfaces. While the single-shot method operates at higher frequencies due to printer size constraints (requiring scaled-down resonators), both platforms demonstrate equivalent physics and successfully realize topological edge states. These approaches can be used in tandem, combining the flexibility of the modular system with the precision and speed of single-shot printing. 

A critical innovation in our approach is the systematic treatment of coupling-induced frequency shifts and the Total Coupling Area (TCA) effect. When acoustic resonators are coupled through tubes, the effective resonant frequency experiences a shift ($\epsilon$ in Eq. \ref{eq:coupling}) that can break the symmetry required for accurate tight-binding implementations. We show how this shift can be mitigated by careful tuning of the Coupling Length (CL), restoring particle-hole symmetry essential for topological phases. Furthermore, we demonstrate that variations in TCA—the total cross-sectional area of all couplings—introduce additional dynamics that can distort the intended band structure. By developing both constant and variable TCA parameterizations, we establish design principles for faithful acoustic implementations of condensed matter models.

To demonstrate the versatility and accuracy of our methods, we realize three increasingly complex topological systems. First, we implement the SSH model, the paradigmatic example of a one-dimensional topological insulator, showing clear bulk-edge correspondence with topological edge modes appearing in the band gap. Second, we construct a Kitaev chain—a minimal model for topological superconductivity. While the Kitaev model has been realized in acoustic systems previously \cite{cavity3}, this represents one of the first acoustic realizations of the Kitaev model with tunable chemical potential. Finally, we study domain boundaries between topological and trivial Kitaev chains, demonstrating that interface modes remain robust and spectrally isolated even when topological and non-topological regions are directly coupled.

The remainder of this paper is organized as follows. Section II introduces the H-resonator design and details both the modular and single-shot fabrication methods. Section III analyzes the dimer system, establishing the principles of positive/negative coupling and the importance of coupling length tuning. Section IV examines the SSH model and reveals the impact of Total Coupling Area on the acoustic band structure. Section V presents the acoustic Kitaev chain implementation, while Section VI demonstrates the topological interface between Kitaev chains with different chemical potentials. Our results establish tube-coupled H-resonator arrays as a flexible and accurate platform for exploring topological physics, with potential applications ranging from acoustic waveguiding to the development of robust acoustic devices based on topological protection.
\begin{figure*}[!t]
    \def\svgwidth{\textwidth}
    \centering
    {\fighel
        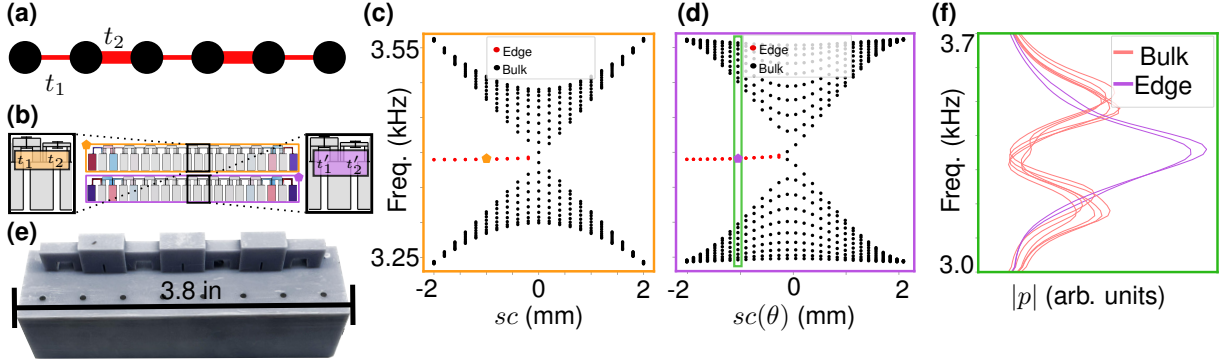
        }
    \caption{ \textbf{SSH schematic, simulation, and experimental measurements.} \textbf{(a)} SSH coupling schematic corresponding to \ref{eq:ssh}. \textbf{(b)} Simulated geometry with acoustic pressure profiles corresponding to the edge modes located at the orange and purple stars in (c) and (d). The geometry in the orange (purple) box corresponds to the non-constant (constant) area parameterization of the coupling widths. \textbf{(c)} SSH band spectrum for the non-constant area coupling parameterization. \textbf{(d)} SSH band spectrum for constant area coupling parameterization. Green box corresponds to the experimentally realized parameters as seen in (e). \textbf{(c,d)} Red dots indicate topological modes localized to the edge as seen in (b). \textbf{(e)} Experimental single-shot SSH system comprised of 8 resonators. \textbf{(f)} Band spectrum for experimental system depicted in (e). Edge modes (purple peaks) centered between bulk peaks (red). }
    \label{fig:ssh-panel}

\end{figure*}

\section{Modular Resonators and One-Shot Printing}
In this work, we present two methods for the in lab manufacturing of acoustic metamaterials: a modular system and one-shot print able solution. Underlying both of these methods is a tube based coupling that allows for effective implementations of tightbinding models in acoustic systems. Results from both of these methods are presented side by side to display different aspects of the underlying tube based coupling system as well as highlight the benefits of each. All plots in the 3 - 4 kHz range correspond to the smaller single-shot method and all plots in the 2.0 - 2.7 kHz range correspond to the modular coupling method. 

Prior to manufacturing, we simulate the internal air volume of the resonator as shown in figure \ref{fig:h-res-design} (a) \& (c) to find its eigenfrequencies. Figure \ref{fig:h-res-design} (b) \& (d) display the first 4 resonant modes of the system and depict the pressure profile of the resonator at each frequency. It is important to select modes that are far enough from the resonant frequency of the microphone and speaker in use to prevent unwanted interference. One can choose any non degenerate resonant mode as the base resonance for hybridization. For discussion about the difference between single and double bridged H-resonators, see Supplement Section \ref{sec:sup1}. In both the modular and single-shot case, we chose mode 1 as the base resonant frequency, outlined in green in figure \ref{fig:h-res-design} (b) (d).

\subsection{Modular Resonators}

The modular platform consists of 3D printed resonator chambers that are open at the top and bottom. These openings are then sealed with interchangeable caps Figure \ref{fig:h-res-design}(b) with ports that allow them to be connected by acrylic tubing Figure \ref{fig:dimer-panel}(c). Both the caps and the base resonators are printed using UV-curable acrylic. To ensure an airtight seal between the caps and resonators, modeling clay is placed at the joint between the cap and the resonator. This  seals the resonator and secures the cap to the chamber. The resonators are then sprayed with soapy water and slightly pressurized while examined for the presence or leakage in the form of bubbles.

\subsection{One Piece Printing}
This second method involves printing the system all at once. To create the 3D-printable geometries, one must simply place a thin shell around the exterior of all geometry elements, design holes to drain the acrylic resin, and place ports for the speakers and microphones in resonators  Figure \ref{fig:h-res-design}(c). In theory, one could manufacture a system as small as the tolerances of the their 3D printer. Given size limitations for our in lab 3D printers, we scale the base resonator by a factor of $1/2$.

\section{Dimer and Negative Coupling}
Once a base resonant mode, $f_o$, has been chosen, one can begin to couple them. To start consider two uncoupled acoustic resonators with resonant frequency $f_o$. They can be represented by the Hamiltonian 
\begin{equation}
    H_R = f_o\ket{R_1} + f_o\ket{R_2}.
\end{equation}

Diagonalizing produces degenerate states with eigenvalue $f_o$. Given the later models use both positive and negative couplings. it is worth discussing two coupled resonators in detail. 
\subsection{Negative Coupling}

 Two H-resonators can be coupled either positively or negatively Figure \ref{fig:dimer-panel} (a, b); thus, the Hamiltonian takes the form 
\begin{equation}
    \label{eq:coupling}
    H_C^{\pm} = (f_o+\epsilon)\ket{R_1}+(f_o+\epsilon)\ket{R_2} \pm v(\ket{R_1} \bra{R_2} + \ket{R_2}\bra{R_1}).
\end{equation}
Where $v$ is the coupling strength that can be varried by adjusting the Coupling Width (CW) Figure \ref{fig:dimer-panel} (c). When coupled positively (negatively) the higher resonant frequency corresponds to the state where the pressure profiles of the two resonators is the same. The epsilon term is due to phase interference introduced by the properties of the coupling. Given that the tight binding model neglects bond interactions, this disanalogizes acoustic systems from atomic systems. Diagonalizing produces eigenvalues of $\lambda_{\pm} = \{f_o+\epsilon \pm v, f_o+\epsilon\mp v$\} and implies the hybridization of the coupled resonators is not symmetric about $f_o$, destroying chiral and particle hole symmetry. However, by tuning the Coupling Lengths (CL) between the resonators Figure \ref{fig:dimer-panel}(c), we can ensure that $\epsilon \to 0$, and  symmetry is preserved Figure \ref{fig:dimer-panel}(d). 

\begin{figure*}[!t]
    \def\svgwidth{\textwidth}
    \centering
    {\fighel
    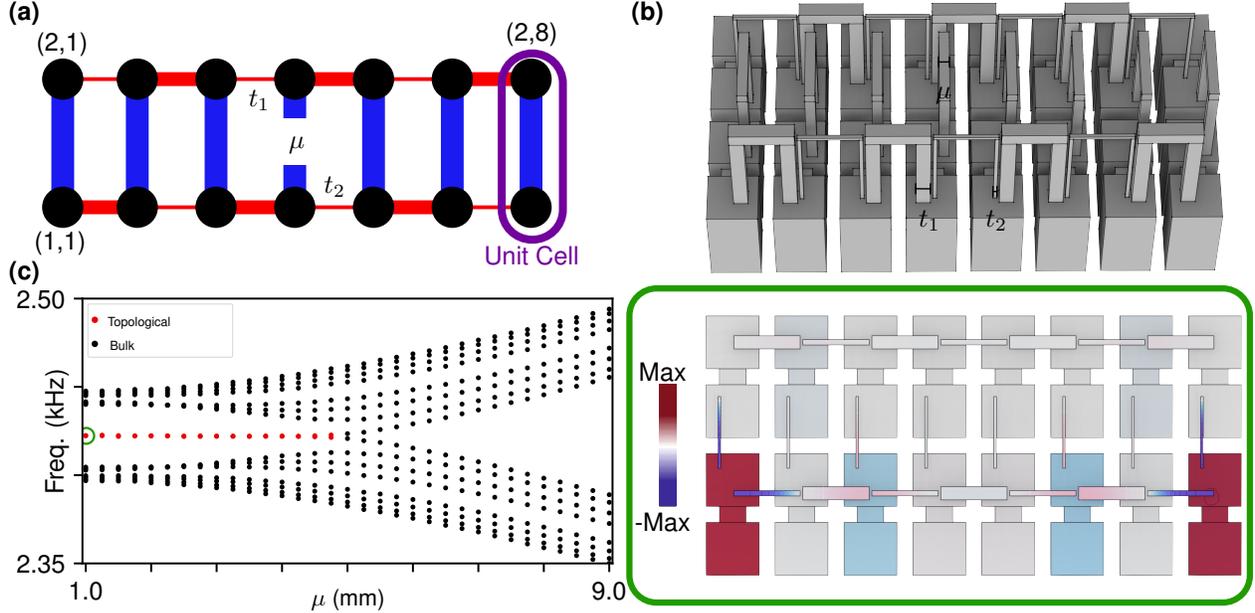
    }
    \caption{\textbf{Kitaev chain schematic, simulation design, and results. }\textbf{(a)} Acoustic Kitaev chain schematic. Thick and think red lines correspond to alternating couplings strengths $t_1$ \& $t_2$. Thick blue line correspond to the onsite hopping potential $\mu$. Unit cell is circled in purple. \textbf{(b)} simulated geometry for the modular method. CW parameters $t_1$, $t_2$, and $\mu$ are labeled. \textbf{(c)} Band spectrum for the Kitaev chain system. Each vertical column of points corresponds to a different $\mu$ value. Topological modes are denoted in red while bulk modes are depicted in black. The mode circled in Green corresponds to the pressure profile depicted in the green box on the right.}
    \label{fig:kitaev}
\end{figure*}

\subsection{Dimer Simulation}

To tune the system, we simulate two resonators in COMSOL Multiphysics coupled by a single tube with length CL and width CW as shown in Figure \ref{fig:dimer-panel} (c). Figure \ref{fig:dimer-panel}(d) shows the simulated eiganfrequencies for three different CLs using the modular system. For each CL, CW is swept from 1 to 9 mm. For the two plots on the right, CW = 90 mm, 128 mm and $\epsilon \ne 0$ and hybridization is not symmetric about $f_o$. For the plot on the left, CW = 109 mm and $\epsilon = 0$, meaning the coupling is accurate for studying tightbinding models. This symmetric dimer plot now acts as a key for deciding coupling strengths. By looking at the frequency as a function of CW, $f(\text{CW})$, we can decide the relative coupling strengths based on the coupling width.

\subsection{Dimer Experiment}

We conduct experimental measurements of the dimer system to corroborate the simulations. The experimental apparatus is shown in figure \ref{fig:dimer-panel} (c) with the modular system on the right and the single shot system on the left. The plots in \ref{fig:dimer-panel} (e) \& (f) were generated by placing a microphone and speaker in the same resonator and playing a pure sinusoidal tone using the audio jack of the computer. We then measure the RMS amplitude of the pressure with the microphone for 0.25 s. The frequency is then increased from 2.2 kHz to 2.5 kHz with a step size of 5 Hz for the modular system and from 3.0 kHz to 4.0 kHz at a step size of 10 for the single-shot system. Figures \ref{fig:dimer-panel} (e) \& (f) show symmetric dimerization of the $f_o$ mode for two experimental setups, each with a different coupling length as expected from our theory and simulations. 

\section{SSH and Total Coupling Area}

\begin{figure*}[!t]
    \def\svgwidth{\textwidth}
    \centering
    {\fighel
        \input{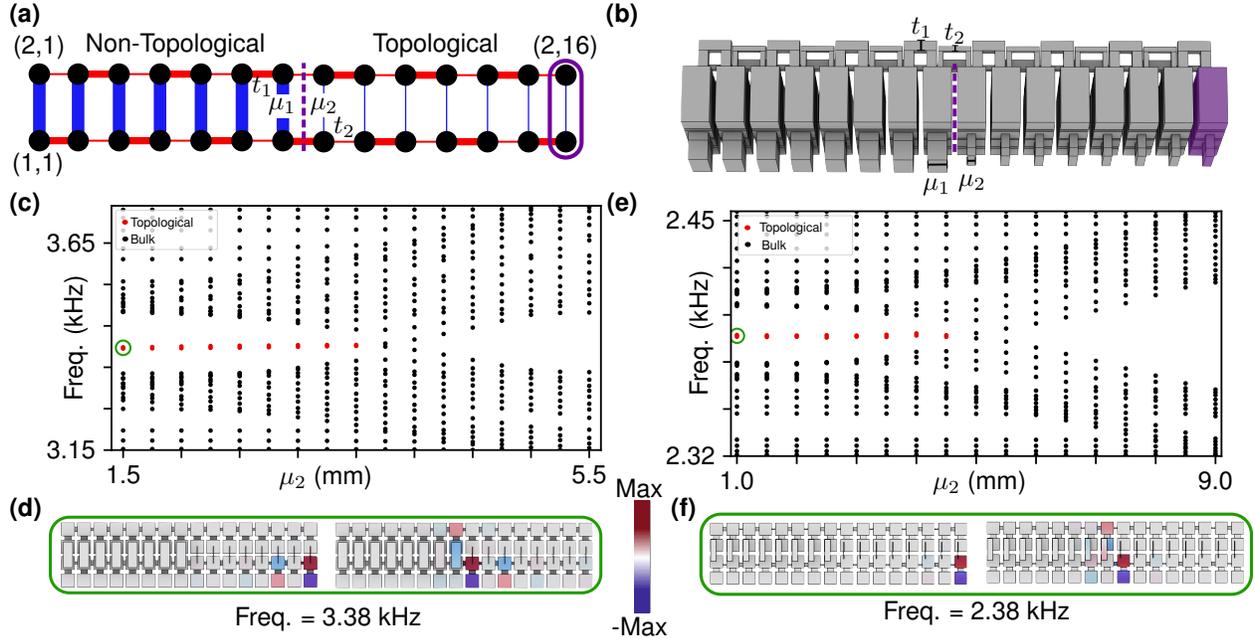}
        }
    \caption{\textbf{Classical Kitaev chain schematic and simulated band spectrum/topological modes.} \textbf{(a)} Single layer Kitaev chain schematic corresponding to real-valued 
    $\Delta$. Coupling strengths $t_1, t_2, \mu_1, \text{and } \mu_2$ are labeled. Positive (negative)  couplings correspond to  Red (blue) lines. \textbf{(b)} Simulated geometry for single-shot Kitaev interface. Coupling widths corresponding to analogous coupling strengths in (a) are labeled. \textbf{(c)} Band spectrum for simulated modular  method with 2 mid gap modes marked in green and red. \textbf{(d)} Edge (interface) corresponding to the green (red) point in (c). \textbf{(e)} Band spectrum for simulated single-shot method with 2 mid gap modes marked in green and red. \textbf{(f)} Edge (interface) mode depicted corresponding to the green (red) point in (e).}
    \label{fig:Simulation-Interface-Panel.svg}
\end{figure*}

Once a CL has been found that ensures $\epsilon \to 0$, we design the whole system so all couplings are CL in length. However additional dynamics are introduced by the variation in the Total Coupling Area (TCA) that are additional to the desired coupling dynamics. The TCA is the total cross sectional area of each different coupling. In many acoustic models of condensed matter systems the dynamics depend on the relative coupling strengths of the system. When the TCA is not constant Figure \ref{fig:ssh-panel}(c), the relative scale of the couplings shifts and additional dynamics are introduced that impact the properties of the system.

To demonstrate this effect, we simulate an SSH system with two different parameterizations of the coupling strengths $t_1$ \& $t_2$ corresponding to the the Hamiltonian 
\begin{equation}
        \label{eq:ssh}
       H_{SSH} = \sum_n [t_1 c_{n,A}^\dagger c_{n,B} + t_2 c_{n,B}^\dagger c_{n+1,A} + h.c.].
\end{equation}\\
Well studied theory tells us that the system goes through a phase transition from trivial to topological insulator as the ratio $t_1/t_2 = r$ increases starting at $r < 1 $ reaching a critical point at $r = 1$, and reaching its topological phase as $r > 1$ \cite{shortcourse}. In our acoustic model the coupling strengths $t_1$ \& $t_2$ are varied by changing the CW Figure \ref{fig:ssh-panel}(a). We define each CW with a common parameter, $sc$ so we can sweep through the different phases of $r$. Letting $t_1 = (4 + sc)$ mm   \& $t_2 = (4 - sc)$ mm, makes $A \propto sc^2$ and as $sc$ increases the total coupling area increases. By sweeping $sc$ we can probe the different regions. We can map the given $sc$ regions to the $r$ regions previously discussed as follows. $sc = 0 \implies r=1$, $sc > 0 \implies r>1$, and $sc<0 \implies r <1$. This creates a warped version of the standard SSH spectrum ass seen in figure \ref{fig:ssh-panel} (c). As the TCA approaches a minima ($sc=0 \text{ \& }  A = 32 \text{ mm}$), the total width of the spectrum shrinks. To demonstrate This effect is due to the area, we define new parameters such that the total coupling area is constant. letting $t_1^{'} = \sqrt{A}\sin{\theta}$ \& $t_2^{'} = \sqrt{A}\cos{\theta}$, the area $A$ is constant due to the pythagorean identity.

 We then sweep theta to create the same regions as the non-constant area model. For ease of comparison, let $sc(\theta) = \frac{t_1^{'}(\theta)-t_2^{'}(\theta)}{2}$ in figure \ref{fig:ssh-panel} (d). When the TCA is fixed the maximum and minimum frequencies in the spectrum remain relatively constant. Notice that the gap remains relatively unaffected between figure \ref{fig:ssh-panel} (c) \& (d). The experimental model for an 8 resonator long SSH chain manufactured using the single-shot method is shown in figure \ref{fig:ssh-panel} (e). The system has fewer resonators than the simulation to accommodate the size of our 3D printer. Despite this, clear topological edge modes are present in the spectral gap shown in the experimental measurements in figure \ref{fig:ssh-panel} (f). 

\subsection{Kitaev Interface Experiment}

\begin{figure*}[!t]
   \def\svgwidth{\textwidth}
   \centering
   {\fighel
        \input{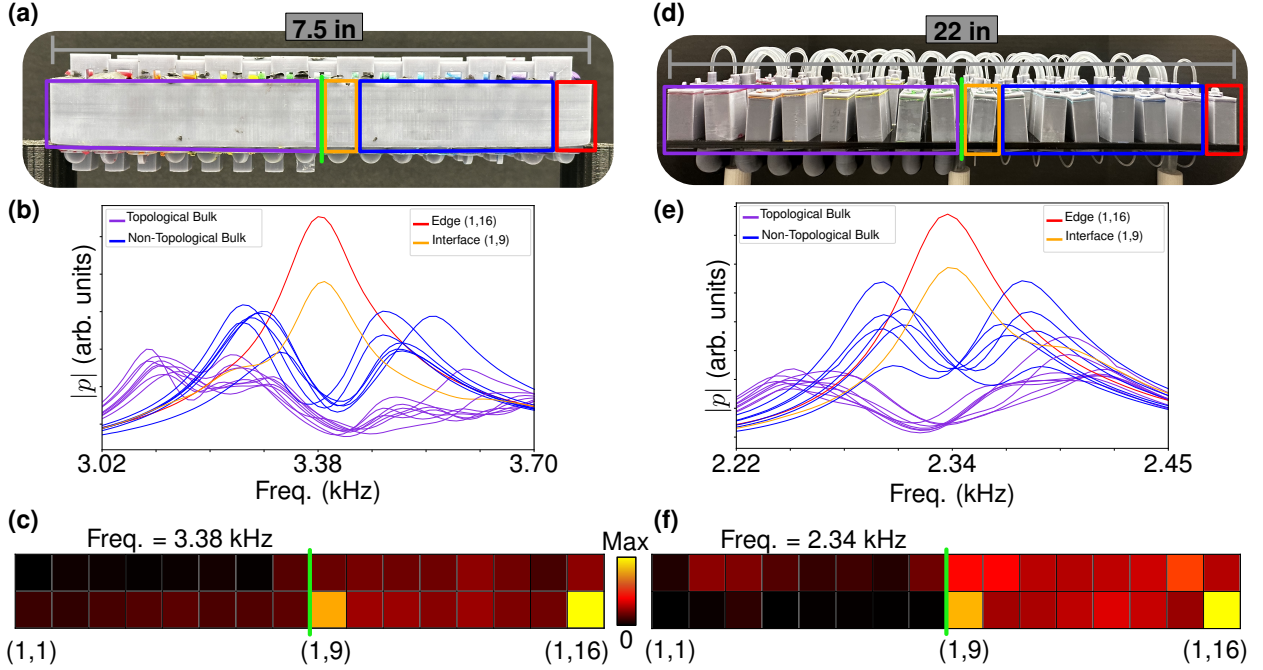}
        }
    \caption{\textbf{Experimental system and measurements for both modular and single shot methods.} \textbf{(a)} Modular Kitaev system. Solid green line depicts the interface between the non-topological (left) and topological (right) Kitaev chains. Blue, orange, purple, and red boxes designate the non-topological bulk, interface site, topological bulk, and edge site  respectively. \textbf{(b)} Experimental measurements for resonators (1,n) are depicted. Colors corresponding to the boxes in (a) designate resonators in different regions. \textbf{(c)} Heat map of normalized pressure intensity for all 32 resonators at the frequency of the edge and interface modes. Pressure legend shown far right. \textbf{(d)} Single-shot interface system. Solid green line depicts the interface between the non-topological (left) and topological (rigt) Kitaev chains. Blue, orange, purple, and red boxes designate the non-topological bulk, interface site, topological bulk, and edge site  respectively. \textbf{(e)} Experimental measurements for resonators (1,n)  are depicted. Colors corresponding to the boxes in (d) designate resonators in different regions. \textbf{(f)} Heat map of normalized pressure intensity for all 32 resonators at the frequency of the edge and interface modes.}
    \label{fig:Expiremental-Interface-Panel.svg}
\end{figure*}

\section{Kitaev Chain}

For some models it is impossible to maintain the total coupling area. Despite this. It is still possible to study aspects of a topological system which we demonstrate using the Kitaev chain. The Kitaev chain provides a minimal theoretical model for one-dimensional topological superconductors supporting Majorana bound states. This model is represented by the Hamiltonian 




\begin{multline}
    H_K = \sum_j -t(c_j^\dagger c_{j+1} + \text{h.c.}) \\
    - \mu(c_j^\dagger c_j - 1/2) + \Delta(c_j c_{j+1} + \text{h.c.})
\end{multline}

where $\Delta = \Delta_x + i \Delta_y$ is the complex order parameter, $\mu$ is the chemical potential, $t$ is the hopping parameter. Given that $\mu$ corresponds to an onsite hoping potential, classical implementations requires a second chain to be placed directly adjacent \cite{spinnerKitaev}. The schematic for the system is displayed in Figure \ref{fig:kitaev} (a). For this model $t_1 = t+\Delta_t$ and $t_2 = t-\Delta_t$. The system exhibits topological properties under the condition that $\mu < \Delta_t$. In this case the total coupling area $A \propto  t_1^2 + t_2^2 +\mu^2$. Given that the dynamics of the system are controlled by the $\mu$ coupling with $t$ \& $\Delta_t$ held constant, it is impossibe to maintain the TCA during the sweep of the mu couplings in Figure \ref{fig:kitaev} (c) and Figure \ref{fig:Simulation-Interface-Panel.svg} (c) \& (e). Despite this, we can still model topological effects. To demonstrate this, we simulate a Kitaev chain of 8 resonators. The simulated model is shown in Figure \ref{fig:kitaev} (b). Figure \ref{fig:kitaev} (c) displays the band spectrum with the midgap topological modes depicted in red. 

\section{Kitaev Chain Interface}

To demonstrate the efficacy of these methods, we employ them to study a domain boundary (interface) between two Kitaev chains, one topological and one non-topological. The coupling schematic is shown it Figure \ref{fig:Simulation-Interface-Panel.svg} (a). Red (blue) lines correspond to positive (negative) couplings. The interface and unit cell for the schematic and geometry are shown as a purple dashed line and box in Figure \ref{fig:Simulation-Interface-Panel.svg} (a) \& (b). To implement this acoustically, we choose two bands corresponding to two different $\mu$ values as seen in figure \ref{fig:kitaev} (c). We choose a $\mu$ value from the right side of the phase shift as the non-topological $\mu_1$ and the topological $\mu_2$ in figure \ref{fig:Simulation-Interface-Panel.svg} (a) \& (b). These two systems are then coupled such that the interface appears as a discontinuous transition between $\mu_1$ \& $\mu_2$ as seen in figure \ref{fig:Simulation-Interface-Panel.svg} (a). These $\mu_1$ \& $\mu_2$ values are chosen to maximize the gap. Thus, we chose the farthest left and farthest right vertical bands.

We simulate this interface with two chains each 8 resonators long. The dimensions of each resonator are the same as those in Figure \ref{fig:h-res-design} (a) \& (c). The simulated geometry for the single shot system is shown in Figure \ref{fig:Simulation-Interface-Panel.svg} (b). $t_1$, $t_2$, and $mu_1$ are held constant at 5.63 mm, 1.41 mm, and 8.89 mm respectively for the modular method and 3.56 mm, 1.81 mm , and 4.5 mm respectively for the single-shot method. These values correspond to the experimental widths of the couplings we use later. $\mu2$ is swept from 1.0 to 9.0 mm in the case of the modular printing Figure \ref{fig:Simulation-Interface-Panel.svg} (e-f) and from 1.5 to 5.5 mm in the case of the single shot printing Figure \ref{fig:Simulation-Interface-Panel.svg} (c-d).  We show that the mid-gap topological modes remain despite contact with a non-topological system for both methods Figure \ref{fig:Simulation-Interface-Panel.svg} (c,e). The degenerate mid-gaped modes are confirmed to be the topological edge and interface modes as shown in figure \ref{fig:Simulation-Interface-Panel.svg} (d) \& (f).

To corroborate these simulations, we perform experimental measurements of the system. We built a system corresponding to a single vertical band in figure \ref{fig:Simulation-Interface-Panel.svg} (c) \& (e). $t_1$, $t_2$, $\mu_1$, $\mu_2$ are  5.63 mm, 1.41 mm, 8.89 mm, and 0.89 mm respectively for the modular method and 3.56 mm, 1.81 mm , 4.5 mm, and 1.5 mm respectively for the single-shot method.  Figure \ref{fig:Expiremental-Interface-Panel.svg} (a) \& (d) depict the finalized experimental systems with left (right) results corresponding to the modular (single-shot) system. measurements were systematically taken for each resonator with their results displayed in Figure \ref{fig:Expiremental-Interface-Panel.svg} (b) \& (c) for the modular system and Figure \ref{fig:Expiremental-Interface-Panel.svg} (e) \& (f) for the single-shot system. The spectrum plots [Figure \ref{fig:Expiremental-Interface-Panel.svg} (b) \& (e)] were generated by plotting the individual resonance spectrum for each of the resonators with in the sweep range. The peaks corresponding to the edge and interface resonators are centered in the gap as expected. The plots in figure \ref{fig:Expiremental-Interface-Panel.svg} (c) \& (d) represent the intensity profile at the peak frequency of the edge mode normalized so that the maximum amplitude is 1. Comparing these to figure \ref{fig:Simulation-Interface-Panel.svg} (d) \& (f) shows that the edge and interface frequencies agree to 3 (2) significant figures in the case of the single-shot (modular) system. The pressure profiles of the individual resonators agree similarly.

\section{Conclusion}

In this work, we have developed two complementary fabrication methods for realizing acoustic analogs of topological tight-binding models using H-shaped resonators with tunable tube couplings. The modular approach offers rapid prototyping and systematic parameter exploration through reconfigurable connections, while the single-shot printing method provides enhanced mechanical stability, reduced acoustic losses, and faster manufacturing for complete metamaterial arrays. Both platforms demonstrate equivalent physics despite operating at different frequency scales, establishing the robustness and versatility of our tube-coupling architecture.

A central achievement of this work is the systematic characterization and mitigation of coupling-induced effects that can distort tight-binding implementations. By tuning the coupling length (CL), we eliminate the frequency shift $\epsilon$ that would otherwise break particle-hole and chiral symmetries essential for accurate topological model realizations. Furthermore, we have identified and quantified the role of Total Coupling Area (TCA) in introducing spurious dynamics, and developed both constant and variable TCA parameterizations to guide faithful acoustic implementations of condensed matter models. These design principles provide a framework for future acoustic metamaterial platforms seeking to implement increasingly complex tight-binding Hamiltonians.

We validated our methods through experimental realization of three topological systems of increasing complexity. The SSH model demonstrated clear bulk-boundary correspondence with robust topological edge states emerging in the spectral gap. The acoustic Kitaev chain—with independently tunable chemical potential—exhibited the expected topological phase transition and Majorana-like edge modes. Most significantly, the interface between topological and non-topological Kitaev chains revealed that interface modes remain spectrally isolated and topologically protected even when distinct phases are directly coupled, confirming the fundamental robustness of topological protection in our platform.

The capability to implement both positive and negative couplings through simple geometric control, combined with the flexibility to access multiple coupling sites on each H-resonator, addresses key limitations that have constrained previous cavity-based acoustic metamaterials. This versatility becomes increasingly important as theoretical models demand more complex coupling architectures that cannot be accommodated by traditional resonator geometries with limited surface area. This paves the way for the experimental realization of more sophisticated systems involving multilayer systems\cite{saunders2025experimentalmilestonesmajoranabraiding}. 

The methods presented here also have practical implications beyond fundamental research. Topologically protected acoustic modes are naturally robust to fabrication imperfections and environmental perturbations, suggesting applications in acoustic waveguiding, sensing, and signal processing devices that require stable operation in noisy environments. The modular system's reconfigurability further enables adaptive acoustic devices whose properties can be dynamically adjusted for specific applications.

In conclusion, tube-coupled H-resonator arrays provide a flexible, accurate, and scalable platform for exploring topological physics in acoustic systems. The combination of systematic design principles, dual fabrication methods, and direct experimental validation establishes this approach as a powerful tool for both fundamental studies of topological phenomena and the development of next-generation acoustic devices leveraging topological protection. As interest in topological materials and classical wave analogs continues to grow, we anticipate that these methods will find broad application in the acoustic metamaterials community and beyond.


\bibliographystyle{naturemag}
\bibliography{refrences}

\clearpage
\appendix
\renewcommand{\thesection}{S\arabic{section}}
\renewcommand{\thesubsection}{S\arabic{section}.\arabic{subsection}}
\renewcommand{\theequation}{S\arabic{equation}}
\renewcommand{\thefigure}{S\arabic{figure}}
\renewcommand{\thetable}{S\arabic{table}}

\setcounter{section}{0}
\setcounter{equation}{0}
\setcounter{figure}{0}
\setcounter{table}{0}

\end{document}